%% file: main.tex
\documentclass{article}
\usepackage{ijcai26}

\usepackage{times}
\usepackage{soul}
\usepackage{url}
\usepackage[hidelinks]{hyperref}
\usepackage[utf8]{inputenc}
\usepackage[small]{caption}
\usepackage{graphicx}
\usepackage{amsmath}
\usepackage{amsthm}
\usepackage{booktabs}
\usepackage{algorithm}
\usepackage{algorithmic}
\usepackage[switch]{lineno}
\usepackage{xspace}   
\usepackage{xcolor}        
\usepackage{amssymb}  
\usepackage{acronym}
\usepackage{verbatim}

\input{macro}

\title{From Secure Agentic AI to Secure Agentic Web: Challenges, Threats, and Future Directions}

\author{
Zhihang Deng$^{1,2}$
\and
Jiaping Gui$^2$\And
Weinan Zhang$^{1,2}$
\affiliations
$^1$Shanghai Innovation Institute\\
$^2$Shanghai Jiao Tong University\\
\emails
\{dzh1227, jgui, wnzhang\}@sjtu.edu.cn
}

\begin{document}

\maketitle

\begin{abstract}
    Large Language Models (LLMs) are increasingly deployed as agentic systems that plan, memorize, and act in open-world environments. This shift brings new security problems: failures are no longer only unsafe text generation, but can become real harm through tool use, persistent memory, and interaction with untrusted web content. In this survey, we provide a transition-oriented view from \emph{Secure Agentic AI} to a \emph{Secure Agentic Web}. We first summarize a component-aligned threat taxonomy covering prompt abuse, environment injection, memory attacks, toolchain abuse, model tampering, and agent network attacks. We then review defense strategies, including prompt hardening, safety-aware decoding, privilege control for tools and APIs, runtime monitoring, continuous red-teaming, and protocol-level security mechanisms. We further discuss how these threats and mitigations escalate in the Agentic Web, where delegation chains, cross-domain interactions, and protocol-mediated ecosystems amplify risks via propagation and composition. Finally, we highlight open challenges for web-scale deployment, such as interoperable identity and authorization, provenance and traceability, ecosystem-level response, and scalable evaluation under adaptive adversaries. Our goal is to connect recent empirical findings with system-level requirements, and to outline practical research directions toward trustworthy agent ecosystems.
\end{abstract}

\input{chapters/01_introduction}

\input{chapters/02_Foundations_of_Agentic_AI_Security}

\input{chapters/03_Threat_Taxonomy}
\input{chapters/04_Defense_Strategies}

\input{chapters/05_Evaluation_and_Benchmarking}
\input{chapters/06_From_Agents_to_the_Agentic_Web}

\input{chapters/07_Open_Challenges_and_Future_Directions}
\input{chapters/08_conclusion}

\section*{Ethical Statement}

There are no ethical issues.

\bibliographystyle{named}
\bibliography{ijcai26}

\end{document}

%% file: macro.tex
\definecolor{red}{RGB}{255,0,0}
\definecolor{green}{RGB}{18,220,168}

\acrodef{wp}[WP]{Website Fingerprinting}

\newcommand{\eat}[1]{}

%% file: chapters/01_introduction.tex
\section{Introduction}
Large Language Models (LLMs) are changing AI systems from just generating text to becoming agentic systems that can plan and act in the real world. By combining an LLM ``brain'' with external tools, memory, and feedback, agents can break down tasks, interact with APIs, browse the web, and carry out multi-step tasks with little human help~\cite{liu2025advances,yu2025survey}. However, this increased freedom also brings new risks: LLM agents are not just models that generate responses but are decision-makers who interact with untrusted environments and real-world systems.

This new autonomy also expands the attack surface. While LLM agents are stochastic and context-sensitive, small adversarial inputs can steer their plans. Prompt injections may directly override objectives, while hidden instructions in webpages, documents, or UI elements can bypass safety rules via indirect injection~\cite{perez2022ignore,greshake2023not}. Attacks may also persist by targeting memory or tool outputs, gradually shifting decisions and eventually triggering harmful actions~\cite{chen2024agentpoison}. Once agents can execute actions (e.g., emails, purchases, device control), these failures can result in real harm.

Consequently, agentic AI security has become a fast-growing topic, with threat models covering injections, tool and protocol risks, and multi-agent manipulation~\cite{datta2025agentic,de2025open}. Defenses span prompt hardening, safer decoding, privilege control, and runtime monitoring~\cite{zhou2024robust,xu2024safedecoding,hu2025agentsentinel,shi2025progent}, supported by benchmarks and red-teaming frameworks~\cite{zhang2024agent,vijayvargiya2025openagentsafety,ge2024mart}. These advances have laid a preliminary foundation for building more reliable and controllable agent systems.

\begin{figure*}[htbp]
\centering
\includegraphics[scale=0.433]{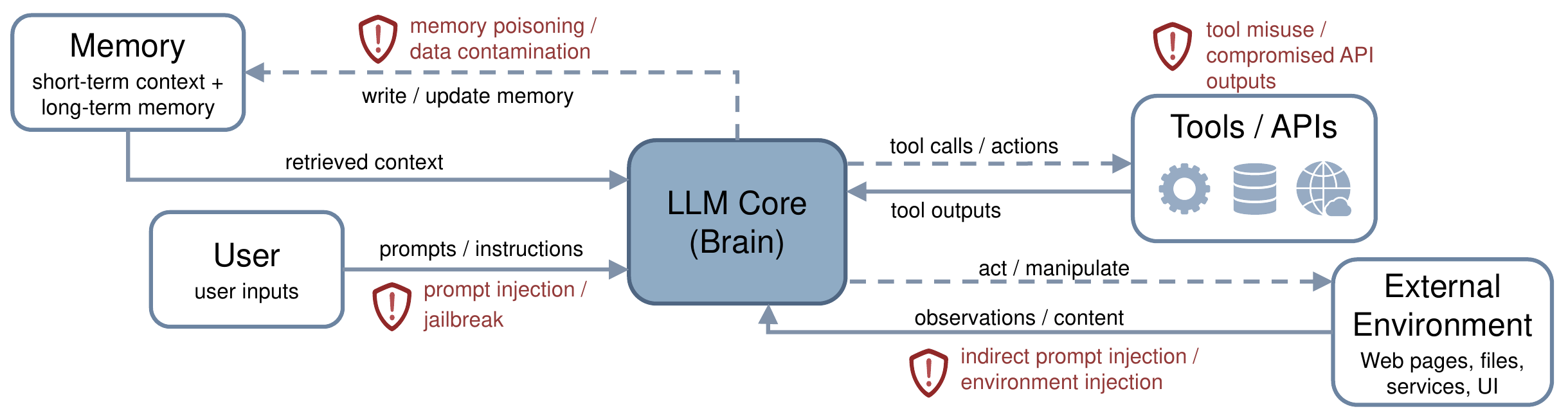}
\caption{An LLM agent architecture with the core model, memory, tool and API interfaces, and environment interaction loop.}
\label{fig:agent_architecture}
\end{figure*}

However, securing individual agents is just part of the problem. A bigger issue is the Agentic Web, the next version of the internet where autonomous agents interact with each other and online services for users~\cite{Yang2025AgenticWW}. This transition changes the security problem in three fundamental ways. First, threats become \emph{networked}: compromises propagate across delegation chains and agent-to-agent communication, amplifying local failures into systemic incidents~\cite{de2025open,lee2024prompt}. Second, trust becomes \emph{programmable}: identity, authorization, and provenance are no longer implicit properties of human users but must be explicit primitives in machine-to-machine interactions, making trust-authorization mismatch a first-class risk~\cite{shi2025sok}. Third, the dominant attack surface shifts toward \emph{tool ecosystems and protocols}: agent actions are mediated by APIs and orchestration layers (e.g., Model Context Protocol), where third-party components and supply-chain-like risks become central~\cite{hou2025model,fang2025we}.

Accordingly, our goal is not to reiterate a general survey of agentic AI security, but to provide a transition-oriented view: we organize threat and defense categories around the agent pipeline and explicitly explain how each category manifests and escalates in the Agentic Web. Concretely, for every threat and defense family (and where relevant, evaluation methodology), we add an ``Agentic Web implication'' paragraph that highlights the shift from single-agent assumptions to open, multi-party, protocol-mediated ecosystems. We then revisit the Agentic Web section with reduced redundancy and focus it on the additional system-level primitives required for web-scale deployment.

%% file: chapters/02_Foundations_of_Agentic_AI_Security.tex
\section{Foundations of Agentic AI Security}
Modern LLM-based agents are modular systems that combine a pretrained LLM with memory, tools/APIs, and an interaction loop with external environments (web pages, files, services, or UIs)~\cite{liu2025advances,yu2025survey}. The Model Context Protocol (MCP) provides a unified abstraction for exchanging context between agents and external tools/services~\cite{hou2025model}. Figure~\ref{fig:agent_architecture} summarizes the key components and information flows.

From a security perspective, Figure~\ref{fig:agent_architecture} is not only a functional diagram, but also a map of attack surfaces. 
Every interface where the agent receives or emits information establishes a trust boundary, each of which can be exploited by malicious actors.
To begin, user inputs may carry malicious prompts aimed at subverting the agent’s original objective, resulting in prompt injection and jailbreak behaviors~\cite{perez2022ignore}.
Secondly, the external environment is usually untrusted and often adversarial by nature.
When agents retrieve and read web pages or documents, concealed instructions within the content can trigger indirect prompt injection, sometimes termed environment injection~\cite{greshake2023not}.
Third, memory components can turn transient attacks into persistent control.
If an attacker manages to write into memory, memory poisoning and data contamination may affect planning decisions long after the initial interaction ends~\cite{chen2024agentpoison}.
Fourth, tools and APIs translate linguistic commands into concrete actions.
Should an agent accept tool outputs without adequate verification, compromised API responses can inject misleading context, and tool misuse may lead to harmful actions being carried out through legitimate interfaces~\cite{li2025stac,fu2024imprompter}.
In essence, the same modules that enable agent autonomy also substantially widen its attack surface; the labels in Figure~\ref{fig:agent_architecture} help to make these risks visually explicit.

Multi‑agent configurations introduce further complexity.
When agents communicate, the output of one becomes the input of another.
This opens new pathways for injection and manipulation across agents, including prompt‑infection‑style propagation and attacks targeting communication channels~\cite{lee2024prompt,he2025red}.
Furthermore, service‑level delegation can blur authorization boundaries.
Risks of trust‑authorization mismatch arise when agents over‑trust peers or services and perform actions beyond their intended scope~\cite{shi2025sok}.
These interactions highlight that security analysis must extend beyond a single agent to consider the integrity of memory systems, toolchains, and communication channels as a whole.

Figure~\ref{fig:taxonomy} abstracts the interfaces in Figure~\ref{fig:agent_architecture} into threat families (Prompt Abuse, Environment Injection, Memory Attacks, Toolchain Abuse, Model Tampering, and Agent Network Attacks) and corresponding defense families (Prompt Hardening, Model Robustness, Tool Control, Runtime Monitoring, Continuous Red Teaming, and Protocol Security). This alignment lets later sections link each paper to a concrete component-level risk and its mitigation.

\paragraph{From Agentic AI security to Agentic Web security.}
While the taxonomy is grounded in Agentic AI security, our survey consistently lifts each category to the Agentic Web setting. In the Agentic Web, agents operate across administrative domains, interact with heterogeneous service agents, and rely on protocol-mediated tool ecosystems~\cite{Yang2025AgenticWW,hou2025model}. This shift introduces new security primitives (e.g., machine identity, delegation constraints, provenance, and cross-domain policy enforcement) and amplifies classical agentic threats via propagation and composition~\cite{de2025open,shi2025sok}. Therefore, Sections~3--5 intentionally pair each agentic AI category with its web-scale manifestation to make the transition explicit.

%% file: chapters/03_Threat_Taxonomy.tex
\section{Threat Taxonomy}
\begin{figure*}[htbp]
    \centering
    \includegraphics[scale=0.31]{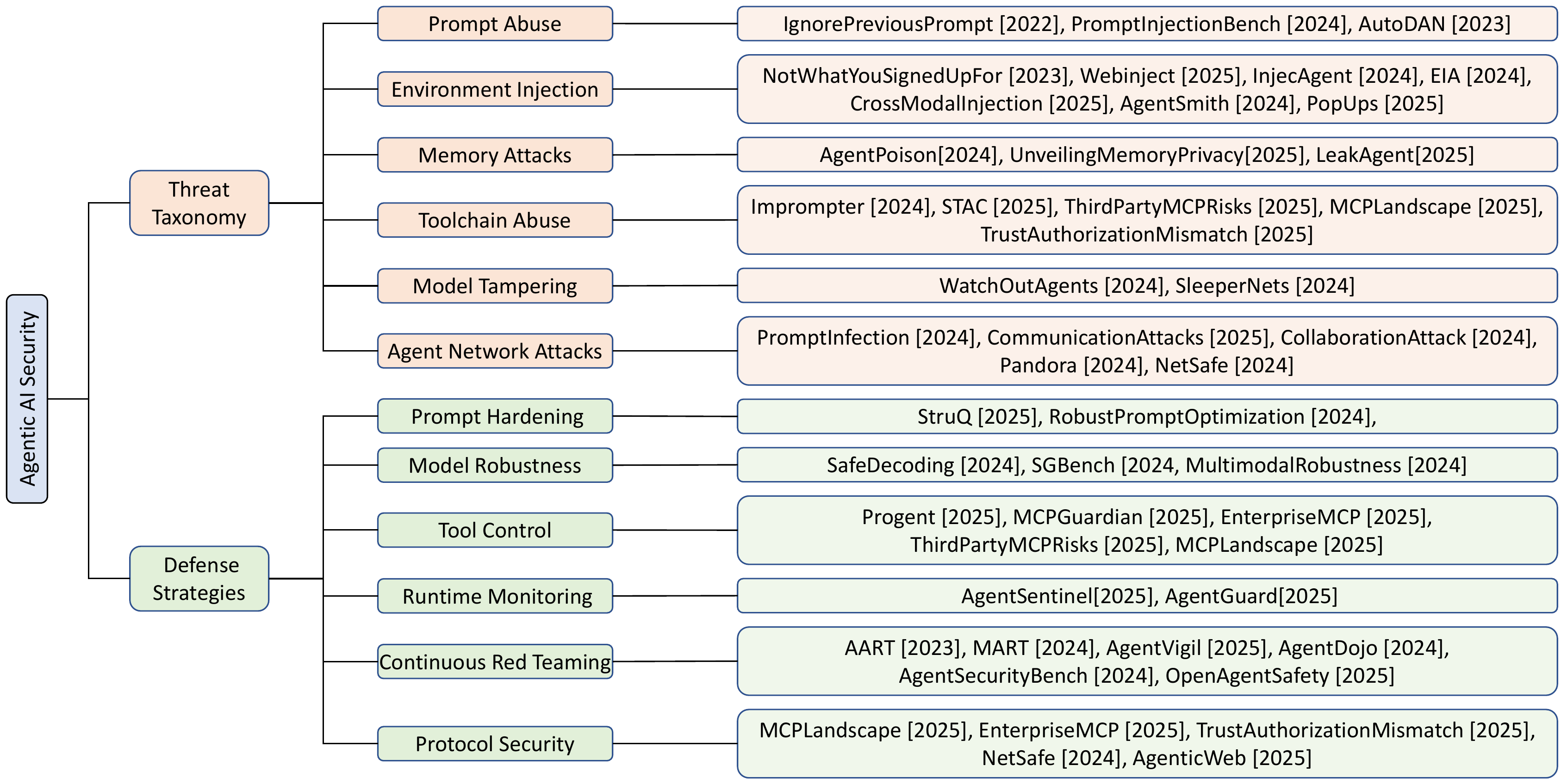}
    \caption{A taxonomy of threats and defenses for agentic AI security aligned with agent components and system layers.}
    \label{fig:taxonomy}
\end{figure*}

This section summarizes core threat classes for agentic AI security with a component oriented view. 
We group threats by where the adversary injects influence: user inputs, external environment, memory, tools and APIs, the model itself, and agent to agent interactions. 
These categories overlap in practice, but they form a clean basis for analysis and for aligning text with Figure~\ref{fig:taxonomy}. 
Surveys such as \cite{yu2025survey,datta2025agentic,deng2025ai} provide complementary threat overviews, while the papers below provide concrete attack mechanisms and measurements.

\subsection{Prompt Abuse}
Prompt abuse targets the agent through user inputs and dialog context. 
The attacker crafts instructions that compete with, override, or reinterpret the agent objective. 
IgnorePreviousPrompt~\cite{perez2022ignore} illustrates that simple directive patterns can bypass instruction hierarchies and safety policies. 
PromptInjectionBench~\cite{liu2024formalizing} shows that prompt injection is not a single trick but a broad family of strategies that generalize across settings, and it provides a benchmarking view that maps well to Figure~\ref{fig:taxonomy}. 
AutoDAN~\cite{liu2023autodan} further indicates that attackers can automatically search for stealthy high success jailbreak prompts against aligned models. 
In agentic settings, prompt abuse becomes more damaging because a compromised plan can be executed through tools and APIs, turning a text level deviation into real actions.

\paragraph{Agentic Web implication.}
In the Agentic Web, prompt abuse can appear inside delegation messages between User Agents and Service Agents, not only in user chat. If the attacker changes the task description or constraints at one hop, the wrong intent can be carried through multi-hop delegation and executed by multiple services via APIs, so the impact is larger than a single-agent jailbreak~\cite{Yang2025AgenticWW,de2025open}.

\subsection{Environment Injection}
Environment injection exploits external environment content that the agent reads, such as web pages, emails, documents, and user interface elements. 
NotWhatYouSignedUpFor~\cite{greshake2023not} demonstrates indirect prompt injection, where malicious instructions are hidden in retrieved content and executed when the agent includes that content in its context. 
InjecAgent~\cite{zhan2024injecagent} operationalizes this risk for tool integrated agents by benchmarking indirect prompt injection cases. 
Webinject~\cite{wang2025webinject} shows how prompt injection can be specialized to web agents that browse and parse untrusted pages. 
EIA~\cite{liao2024eia} highlights a privacy focused variant, where the environment is crafted to induce unintended disclosure by generalist web agents. 
In multimodal agents, CrossModalInjection~\cite{wang2025manipulating} expands the attack surface by embedding malicious instructions across modalities. 
AgentSmith~\cite{gu2024agent} shows an extreme amplification pattern, where a single image can trigger large scale multimodal agent jailbreaking. 
PopUpAttack~\cite{zhang2025attacking} illustrates that user interface elements such as pop ups can steer computer use agents toward unsafe clicks or disclosures. 
Overall, environment injection is central for secure agentic web settings because the web is an untrusted and adversarial environment by default.

\paragraph{Agentic Web implication.}
In the Agentic Web, agents continuously read untrusted web content, so indirect injection becomes a common and scalable attack path~\cite{Yang2025AgenticWW,wang2025webinject}. A malicious page can affect not only the agent that reads it, but also other agents that consume its summaries, extracted data, or delegated subtasks, making one injection spread across a workflow~\cite{lee2024prompt,de2025open}.

\subsection{Memory Attacks}
Memory attacks target the memory components of AI agents, deliberately corrupting or manipulating this persistent data to subvert the agent's future reasoning. 
AgentPoison~\cite{chen2024agentpoison} shows that poisoning memory or knowledge bases can influence future reasoning and planning toward an attacker goal. 
MemoryPrivacyRisks~\cite{wang2025unveiling} shows that when memory stores sensitive user data, the agent can be induced to disclose it intentionally or accidentally. 
LeakAgent~\cite{nie2025leakagent} further shows that privacy leakage can be systematically elicited and optimized by a learning based red teaming agent, rather than discovered by chance. 
Memory attacks are especially important for long horizon agents, where persistent context is a core capability and also a persistent liability.

\paragraph{Agentic Web implication.}
In the Agentic Web, memory is often shared across sessions, users, or services (e.g., org knowledge bases or retrieval backends), so poisoning can persist and be reused many times~\cite{chen2024agentpoison,wang2025unveiling}. This turns a one-time write into long-term influence on future tasks. Privacy leakage also scales because long-horizon web agents may accumulate traces from many delegated steps, and attackers can systematically trigger disclosure~\cite{nie2025leakagent}.

\subsection{Toolchain Abuse}
Toolchain abuse targets tools and APIs, including tool outputs, tool routing, and chains of tool calls. 
Imprompter~\cite{fu2024imprompter} shows that attackers can trick agents into improper tool use, causing unsafe tool calls even when the agent appears aligned at the text level. 
STAC~\cite{li2025stac} shows that seemingly innocent tools can form dangerous chains, producing jailbreak effects through composition rather than a single prompt. 
At a broader system layer, ThirdPartyMCPRisks~\cite{fang2025we} argues that third party components in MCP powered systems introduce safety risks that resemble supply chain problems. 
MCPLandscape~\cite{hou2025model} provides a structured analysis of threats around the Model Context Protocol, which directly connects to tool mediated context exchange. 
TrustAuthorizationMismatch~\cite{shi2025sok} highlights a recurring failure mode where trust signals and authorization boundaries do not match, leading to unintended capabilities in agent service interactions. 
These threats connect directly to secure agentic web deployment where tool mediated actions are the main path from language to impact.

\paragraph{Agentic Web implication.}
In the Agentic Web, most real impact comes from tool and API calls, and agents rely on many third-party services and orchestration layers (e.g., MCP)~\cite{hou2025model,fang2025we}. This makes toolchain abuse closer to a supply-chain problem: a compromised tool/provider or a bad tool combination can cause harmful actions even if the model output looks harmless~\cite{li2025stac}.

\subsection{Model Tampering}
Model tampering targets the model or policy itself, including backdoors and hidden triggers. 
WatchOutAgents~\cite{yang2024watch} shows that an LLM based agent can behave normally under standard tests while failing catastrophically under trigger conditions. 
SleeperNets~\cite{rathbun2024sleepernets} shows that backdoor poisoning can create triggers that remain dormant until specific deployment states appear, which complicates detection and auditing.  
Model tampering is difficult to mitigate purely at runtime because the failure mode is conditional and can evade detection until activation.

\paragraph{Agentic Web implication.}
In the Agentic Web, users may interact with many Service Agents that run different models, and agents may dynamically select models or plugins, so the model supply chain becomes part of the threat surface~\cite{hou2025model}. Backdoors are especially risky because triggers may only appear in specific real workflows, which are hard to cover in tests~\cite{yang2024watch,rathbun2024sleepernets}.

\subsection{Agent Network Attacks}
Agent network attacks emerge in settings where multiple agents interact, leading to situations where the output of one agent becomes the input for another.
PromptInfection~\cite{lee2024prompt} shows that LLM to LLM communication can carry injection payloads across agents, which aligns with the Agent Network Attacks branch in Figure~\ref{fig:taxonomy}. 
In a similar direction, CommAttackRedTeam~\cite{he2025red} offers a systematic set of communication attack strategies aimed at red‑teaming multi‑agent systems.
MultiagentCollaborationAttack~\cite{amayuelas2024multiagent} shows that debate style collaboration can be exploited to steer agents toward unsafe outcomes. 
Pandora~\cite{chen2024pandora} demonstrates how collaborative phishing agents can employ decomposed reasoning to achieve detailed jailbreaking.
Finally, NetSafe~\cite{yu2024netsafe} argues that overall safety depends not merely on the robustness of individual agents, but also on the underlying network structure and the influence pathways that connect agents.

\paragraph{Agentic Web implication.}
In the Agentic Web, agent‑to‑agent communication forms the fundamental infrastructure for task execution. This introduces a systemic security vulnerability: a single compromised agent can influence many others through messages and delegated subtasks~\cite{Yang2025AgenticWW,lee2024prompt}. This threat can propagate covertly across the collaboration graph, triggering cascading failures that are difficult to trace, with an impact far exceeding that of an isolated agent failure. Consequently, security considerations must shift from ensuring single-agent robustness to modeling and analyzing the pathways of trust propagation and risk contagion across the entire networked ecosystem~\cite{he2025red,yu2024netsafe}.

%% file: chapters/04_Defense_Strategies.tex
\section{Defense Strategies}
This section summarizes major defensive approaches for agentic AI security using a layered design view. 
We organize defenses according to where they act in the agent pipeline: prompt and context construction, model behavior control, tool and APIs control, runtime monitoring, evaluation and red teaming, and protocol  trust mechanisms. 
It is worth noting that these categories are complementary, and practical systems typically combine them into defense in depth \cite{yu2025survey,datta2025agentic}.

\subsection{Prompt Hardening}
Prompt hardening improves the separation between trusted instructions and untrusted content. 
StruQ~\cite{chen2025struq} uses structured queries to prevent user supplied text from being interpreted as executable instructions, reducing injection success in tool integrated agents. 
RobustPromptOpt~\cite{zhou2024robust} searches for prompts that are harder to jailbreak and can be used as a system level guardrail. 
Prompt hardening is necessary but not sufficient because Environment Injection and Toolchain Abuse can still introduce adversarial content through non user channels.

\paragraph{Agentic Web implication.}
In the Agentic Web, prompt hardening must cover not only user inputs but also delegation messages and service responses. A practical goal is to make message formats explicit (what is instruction vs. what is content) and enforce this separation at the orchestration/protocol boundary, because otherwise a single hop can reintroduce hidden instructions into the chain~\cite{Yang2025AgenticWW,hou2025model}.

\subsection{Model Robustness}
Model robustness controls model behavior even under adversarial pressure. 
SafeDecoding~\cite{xu2024safedecoding} constrains generation so that unsafe continuations are less likely, even when the prompt is adversarial. 
SGBench~\cite{mou2024sg} shows that robust safety behavior must generalize across diverse tasks and prompt types, not only within a narrow test set. 
MultiModalRobustness~\cite{wu2024dissecting} shows that robustness must cover both language and perception channels for multimodal agents. 
These approaches reduce the chance that the model produces unsafe plans, but they do not by themselves enforce tool level safety.

\paragraph{Agentic Web implication.}
In the Agentic Web, models face much more diverse and adversarial conditions (pop-ups, long contexts, multimodal pages), so robustness must generalize beyond a fixed prompt set~\cite{wu2024dissecting,zhang2025attacking}. Robustness evaluation should therefore include interactive web workflows and multi-step traces, not only single-turn prompts~\cite{mou2024sg}.

\subsection{Tool Control}
Tool control limits what actions an agent can take through tools and APIs. 
ProGENT~\cite{shi2025progent} enforces programmable privilege control at the tool interface, blocking calls outside an allowed policy. 
MCPGuardian~\cite{kumar2025mcp} proposes a security first mediation layer that validates tool requests and tool outputs in MCP based systems. 
EnterpriseMCP~\cite{narajala2025enterprise} extends this direction with enterprise grade security frameworks and mitigation strategies for MCP deployments. 
ThirdPartyMCPRisk~\cite{fang2025we} and MCPSecurityLandscape~\cite{hou2025model} both motivate explicit trust boundaries and verification for tool providers and external services. 
Tool control is central for secure agentic web settings because real world impact is usually mediated by APIs.

\paragraph{Agentic Web implication.}
In the Agentic Web, safety is largely determined by what tools and APIs an agent is allowed to call, since most real impact is mediated by services~\cite{shi2025progent}. This makes least-privilege permissions, delegation constraints, and tool mediation at protocol gateways especially important, including handling third-party tool risk in MCP-like ecosystems~\cite{hou2025model,kumar2025mcp,fang2025we}.

\subsection{Runtime Monitoring}
Runtime monitoring adds online oversight for agent plans and actions. 
AgentSentinel~\cite{hu2025agentsentinel} provides an end to end real time defense framework for computer use agents by monitoring action streams and enabling blocking or rollback. 
AgentGuard~\cite{chen2025agentguard} repurposes an agentic orchestrator for safety evaluation of tool orchestration, and its evaluation signals can be used as monitoring cues for unsafe tool usage patterns. 
Runtime monitoring is especially valuable when threats adapt to bypass static prompt hardening.

\paragraph{Agentic Web implication.}
In the Agentic Web, monitoring should track end-to-end delegation chains rather than only one agent’s local actions, because failures can propagate across agents and services~\cite{de2025open}. Monitoring should support tracing who delegated what and allow interventions like blocking, rollback, or quarantining suspicious agents/services~\cite{hu2025agentsentinel,chen2025agentguard}.

\subsection{Continuous Red Teaming}
Continuous red teaming treats agent security as an iterative process, not a one time audit. 
AART~\cite{radharapu2023aart} generates diverse adversarial cases for new LLM powered applications, supporting continuous stress testing. 
MART~\cite{ge2024mart} improves safety with multi round automatic red teaming by repeatedly discovering failures and updating defenses. 
AgentVigil~\cite{wang2025agentvigil} provides generic black box red teaming for indirect prompt injection against agents, matching the Environment Injection branch in Figure~\ref{fig:taxonomy}. 
AgentDojo~\cite{debenedetti2024agentdojo} offers a dynamic environment to evaluate prompt injection attacks and defenses for LLM agents. 
AgentSecurityBench~\cite{zhang2024agent} formalizes and benchmarks attacks and defenses in LLM based agents, making it useful as a stable evaluation backbone. 
OpenAgentSafety~\cite{vijayvargiya2025openagentsafety} evaluates real world agent safety across tasks and deployment conditions.

\subsection{Protocol Security}
Protocol security protects agent to service and agent to agent interactions through identity, authorization, and trust controls. 
MCPSecurityLandscape~\cite{hou2025model} and EnterpriseMCP~\cite{narajala2025enterprise} highlight protocol specific threats and mitigation directions for tool mediated context exchange. 
TrustAuthorizationMismatch~\cite{shi2025sok} motivates explicit authorization checks and safer delegation policies in agent interactions. 
AgenticWeb~\cite{Yang2025AgenticWW} motivates standardized secure interaction layers so that large agent ecosystems can interoperate without implicit trust.

\paragraph{Agentic Web implication.}
In the Agentic Web, protocol security is what makes cross-domain delegation safe: agents need reliable identity, authorization, and provenance checks across services, plus revocation when something is compromised~\cite{Yang2025AgenticWW,shi2025sok}. MCP-focused analyses show that without clear trust boundaries and third-party controls, protocol-level context exchange can become a high-risk choke point~\cite{hou2025model,narajala2025enterprise}.

%% file: chapters/05_Evaluation_and_Benchmarking.tex
\section{Evaluation and Benchmarking}
Rigorous evaluation is essential for secure agent development: it turns threat models into repeatable tests, supports fair comparison of defenses, and helps detect safety regressions as agent designs evolve. Recent work has therefore moved from ad-hoc demos toward benchmarks, end-to-end frameworks, and automated red-teaming that better reflect tool-integrated agent behavior.

Standardized benchmarks provide controlled testbeds for measuring attack success and defense effectiveness. Agent Security Bench (ASB)~\cite{zhang2024agent} formalizes common agent attack surfaces such as prompt injection and improper tool use, enabling reproducible evaluation and systematic ablations. Complementing this, end-to-end frameworks aim to approximate real deployments. OpenAgentSafety~\cite{vijayvargiya2025openagentsafety} evaluates agents across diverse tasks and environments with metrics tied to instruction-following, policy compliance, and robustness to malicious inputs. AgentDojo~\cite{debenedetti2024agentdojo} focuses on prompt injection in a dynamic interactive environment, making it useful for studying when agents drift from intended behavior under injected content and for testing prompt-hardening defenses.

A complementary direction treats automated red-teaming as evaluation infrastructure. AART~\cite{radharapu2023aart} generates diverse adversarial data to continuously probe new LLM applications, while MART~\cite{ge2024mart} runs multi-round adversarial interactions to expose weaknesses and track improvements over iterations. In parallel, specialized evaluations target high-stakes failure modes such as privacy: LeakAgent~\cite{nie2025leakagent} uses an RL-based attacker agent to quantify privacy leakage risk, and memory-focused studies quantify how long-term memory can inadvertently reveal sensitive content~\cite{wang2025unveiling}. Finally, as agent systems scale to multi-agent settings, evaluation is extending to system-level trade-offs. The ``multi-agent security tax''~\cite{peigne2025multi} quantifies how security measures can reduce collaboration capability, and broader discussions call for large-scale testbeds to study emergent risks like collusion and cascades~\cite{de2025open}.

In summary, the evaluation landscape is converging on a practical toolkit: controlled benchmarks for reproducibility, end-to-end frameworks for realism, and automated or targeted assessments for continuously evolving threats.

%% file: chapters/06_From_Agents_to_the_Agentic_Web.tex
\section{From Agents to the Agentic Web}

\begin{figure}[htbp]
    \centering
    \includegraphics[scale=0.32]{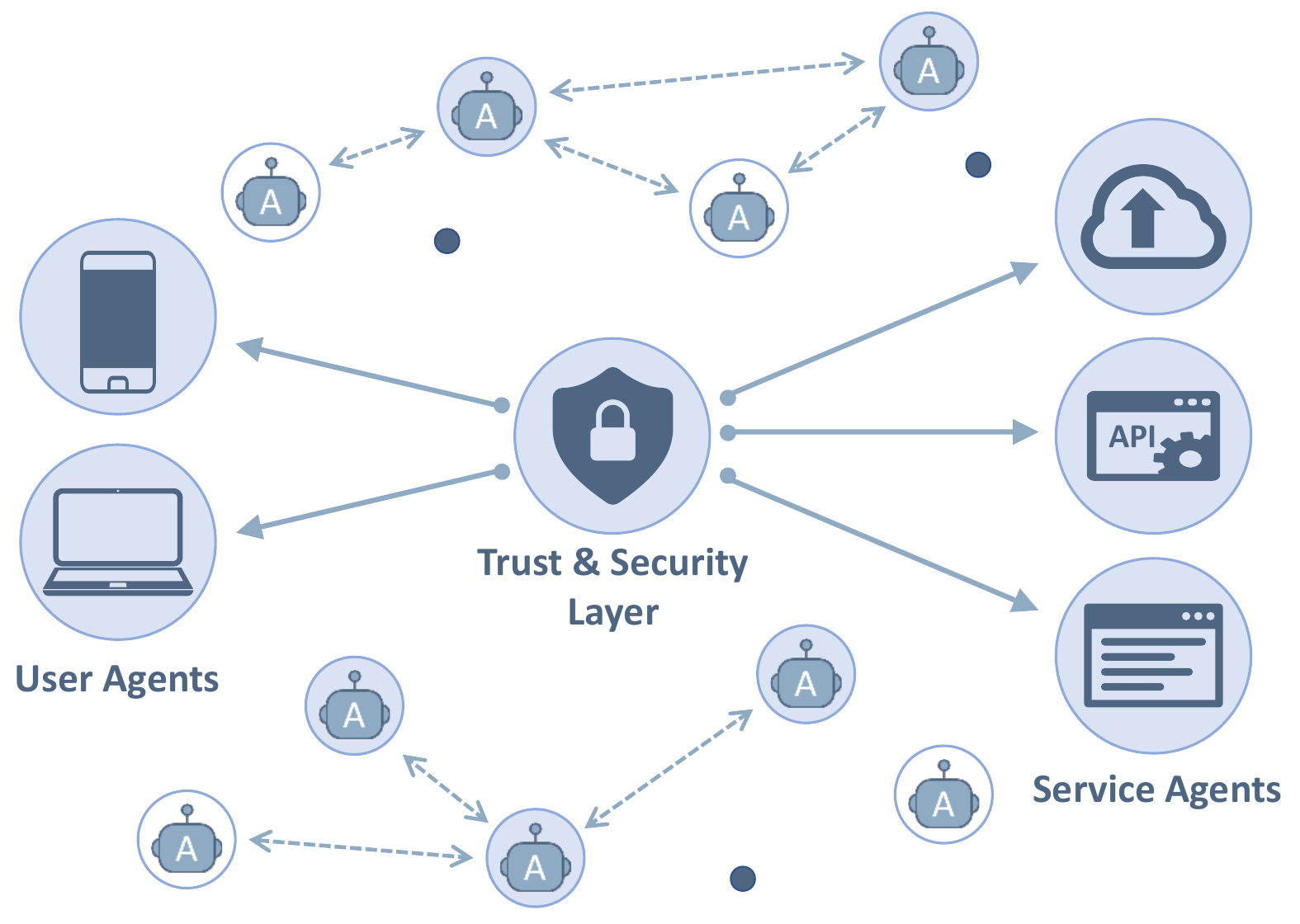}
    \caption{An illustration of the Agentic Web concept.}
    \label{fig:Agentic_Web}
\end{figure}

Earlier sections mapped the main threat families in agentic systems, reviewed layered defenses, and summarized how evaluation and red teaming are becoming practical infrastructure for secure development.These threads point to the same conclusion. In agentic systems, the security boundary is no longer the perimeter of a single agent. It becomes the perimeter of an ecosystem. Failures can be amplified when tools and services are chained together, and when messages and delegated tasks move between agents and services.This section brings these observations together and explains what must be added when moving from Secure Agentic AI to a Secure Agentic Web.

The Agentic Web envisions an internet where autonomous agents discover, communicate, and transact on behalf of users, shifting the web from a human-centric information space to an agent-centric action network~\cite{Yang2025AgenticWW}. 
Compared with standalone deployments, web-scale agentic systems are characterized by (i) cross-domain delegation chains, (ii) heterogeneous service agents and tool ecosystems, and (iii) protocol-mediated interoperability. 
As illustrated in Fig.~\ref{fig:Agentic_Web}, these properties motivate a separation of roles between \emph{User Agents} and \emph{Service Agents}, together with a \emph{Trust \& Security Layer} that makes identity, authorization, and policy enforcement explicit and checkable rather than implicit assumptions.

We therefore emphasize three additional primitives required beyond single-agent security.
\textbf{(1) Interoperable identity and authorization:} agents and services need machine-verifiable identities and explicit delegation constraints, addressing trust-authorization mismatch as a first-class failure mode~\cite{shi2025sok}. 
\textbf{(2) Provenance and traceability:} delegation requests, intermediate artifacts (summaries, plans), and tool outputs should carry provenance metadata and be auditable, enabling both prevention (rejecting untrusted sources) and post-incident forensics. 
\textbf{(3) Ecosystem-level response:} web-scale deployments require operational mechanisms such as quarantine, revocation, and recovery to limit blast radius when open-world attacks inevitably occur~\cite{de2025open}. 
These primitives reframe agentic web security as an ecosystem governance problem with enforceable technical controls, complementing the Agentic AI Security.

%% file: chapters/07_Open_Challenges_and_Future_Directions.tex
\section{Open Challenges and Future Directions}
Despite recent progress, secure agentic AI remains in its infancy, and many open problems must be addressed to safely scale up to an agentic web. Below we distill the most pressing challenges and directions into five themes.

\begin{itemize}
    \item \textbf{Scaling security from single agents to agent societies:} Current defense techniques are often validated on individual agents or small-scale scenarios, but their effectiveness at web scale remains unclear. As agent ecosystems grow, security failures can become systemic rather than local. New methodologies are needed to manage the combinatorial complexity of securing large agent ecosystems~\cite{kim2025towards,de2025open}. This includes security infrastructures that can monitor and coordinate many agents without becoming bottlenecks, and approaches that can detect emergent threats that may not be visible when examining agents in isolation.

    \item \textbf{Trust, identity, and authorization as first-class primitives:} Establishing trust in an agentic web is an interdisciplinary problem spanning cryptography, reputation systems, and governance. Agents need mechanisms to authenticate peers, reason about trustworthiness, and verify that delegated actions remain within authorized scopes. A core difficulty is trust authorization mismatch, where an agent over-trusts a counterpart or over-grants permissions, creating an exploitable gap~\cite{shi2025sok}. Practical directions include robust identity infrastructures for agents, provenance tracking of plans and tool outputs, and delegation protocols with explicit constraints that can be checked and enforced.

    \item \textbf{End-to-end safety for tool use and delegation chains:} Agentic systems rely on tool orchestration and multi-hop delegation, which introduces new failure modes when benign components form dangerous chains. Work on privilege control and the ``security tax'' highlights a fundamental tension: tighter controls reduce risk but can constrain capability and usability~\cite{shi2025progent,peigne2025multi}. Future research should develop end-to-end designs that combine least privilege, runtime monitoring, and safe composition of tools so that delegation chains remain safe even under partial compromise and adversarial influence.

    \item \textbf{Robustness and assurance beyond patchwork defenses:} Many defenses today are narrow and fragile, addressing particular attack classes while offering limited guarantees outside their intended scope. A major direction is to improve generalizable robustness and move toward stronger assurance, including formal specifications of forbidden actions, verifiable constraints over tool calls, and systematic reasoning audits. While full formal verification of LLM-driven autonomy is difficult, intermediate goals such as constrained action spaces, checkable policies, and validated reasoning steps can provide practical safety gains at scale~\cite{datta2025agentic,yu2025survey}.

    \item \textbf{Continuous evaluation and adaptive security against evolving adversaries:} Agentic security will remain a moving target as attackers adapt. This motivates continuous red-teaming, standardized evaluation, and rapid mitigation workflows. Recent work underscores the value of automated and real-world oriented safety evaluation for agents~\cite{vijayvargiya2025openagentsafety,zhang2024agent}. Another emerging direction is antifragility, where systems improve through exposure to stressors and attacks, provided the adaptation process itself is safe and well-governed~\cite{jin2025position}.
\end{itemize}

Taken together, these directions suggest that a secure agentic web requires both foundational primitives and engineering discipline. Additional concerns such as cross-domain and physical-world threats in embodied and cyber-physical deployments~\cite{eslami2025security}, third-party and protocol level risks in agent infrastructures and ecosystems~\cite{fang2025we,hou2025model}, and standardization and policy coordination will also be essential. Progress will likely come from tightly coupling technical advances in security, evaluation, and oversight with interoperable standards that make trust and safety enforceable by design.

%% file: chapters/08_conclusion.tex
\section{Conclusion}
The security of Agentic AI is a rapidly evolving challenge. The introduction of new capabilities in autonomous agents also brings novel vulnerabilities that require constant attention. This survey has mapped the landscape of threats, from prompt-level attacks to complex multi-agent exploits, and reviewed defense strategies across multiple layers of security, including robust prompting, tool control, and runtime monitoring. A clear conclusion from this work is that no single security measure will suffice; a layered, defense-in-depth approach is crucial for ensuring the robustness of autonomous agents. Moreover, the transition from today’s individual agents to a future agentic web amplifies these security challenges, making it imperative to establish a solid security foundation now.

As we move towards the realization of the Agentic Web, the need for secure-by-design frameworks becomes even more urgent. By drawing from the latest research and technological innovations, we can inspire new approaches to secure agent ecosystems. Collaborations between AI researchers, security experts, and policymakers will be key in addressing these challenges. Through sustained effort, we can realize the full potential of Agentic AI to drive productivity and foster new services, while maintaining safety and trust. Achieving a secure Agentic Web is an ambitious goal, but with the community's concerted efforts, it is one that is within reach.